# An optimal security management framework for backhaul-aware 5G-Vehicle to Everything (V2X)

*Vishal Sharma, Jiyoon Kim, Yongho Ko, Ilsun You\*, Jung Taek Seo*
*Information Security Engineering Department,*
*Soonchunhyang University, Asan-si-31538,*
*The Republic of Korea*
vishal_sharma2012@hotmail.com, {74jykim, koyh0911, ilsunu}@gmail.com, seojt@sch.ac.kr

## Abstract

Cellular (C) setups facilitate the connectivity amongst the devices with better provisioning of services to its users. Vehicular networks are one of the representative setups that aim at expanding their functionalities by using the available cellular systems like Long Term Evolution (LTE)-based Evolved Universal Terrestrial Radio Access Network (E-UTRAN) as well as the upcoming Fifth Generation (5G)-based functional architecture. The vehicular networks include Vehicle to Vehicle (V2V), Vehicle to Infrastructure (V2I), Vehicle to Pedestrian (V2P) and Vehicle to Network (V2N), all of which are referred to as Vehicle to Everything (V2X). 5G has dominated the vehicular network and most of the upcoming research is motivated towards the fully functional utilization of 5G-V2X. Despite that, credential management and edge-initiated security are yet to be resolved under 5G-V2X. To further understand the issue, this paper presents security management as a principle of sustainability and key-management. The performance tradeoff is evaluated with the key-updates required to maintain a secure connection between the vehicles and the 5G-terminals. The proposed approach aims at the utilization of high-speed mmWave-based backhaul for enhancing the security operations between the core and the sub-divided functions at the edge of the network through a dual security management framework. The evaluations are conducted using numerical simulations, which help to understand the impact on the sustainability of connections as well as identification of the fail-safe points for secure and fast operations. Furthermore, the evaluations help to follow the multiple tradeoffs of security and performance based on the metrics like mandatory key updates, the range of operations and the probability of connectivity.

## 1 Introduction

The new radio access technology has led to a major transition from Long Term Evolution (LTE)-based Evolved Universal Terrestrial Radio Access Network (E-UTRAN) to Fifth Generation (5G)-based functional architecture for vehicular networks. Such vehicular networks, which are predominately supported by the Cellular (C) setups, are termed as C-Vehicle to Everything (C-V2X). Enhancing applications with a range of operations of C-V2X are one of the key advantages of 5G networks [1] [2] [3].

5G-V2X involves a multi-modular architecture where several functions are considered for securing the services for vehicles involved in the formation of the Vehicle to Vehicle (V2V), Vehicle to Infrastructure (V2I), Vehicle to Pedestrian (V2P) and Vehicle to Network (V2N) as a part of V2X, as shown in Fig.1. These networks require a special implementation that provides strong fronthaul connectivity and edge-initiated security which can simultaneously manage the backhaul operations [3] [4].

Enhancing networks towards edge-infrastructure and bringing security functions near to the users are tedious tasks to be attained under 5G-V2X. Many studies have suggested implementing slice-based V2X, however, a sizably voluminous number of key updates and non-availability of edge-functions lead to overheads and deteriorate the performance. Additionally, with most of the functions operable at the core, the distance of operations increases, and it alarms the possibilities of cyber-attacks on both vehicles and connected components.

There are certain studies that have presented security management as a resource allocation problem and proposed an optimized network as a solution for enhancing the performance of V2X [5] [6] [14]-[24]. These studies do not consider any strategic management for combining the security of backhaul formed between the Terminal (TM) and the hub with the edge-initiated operations.

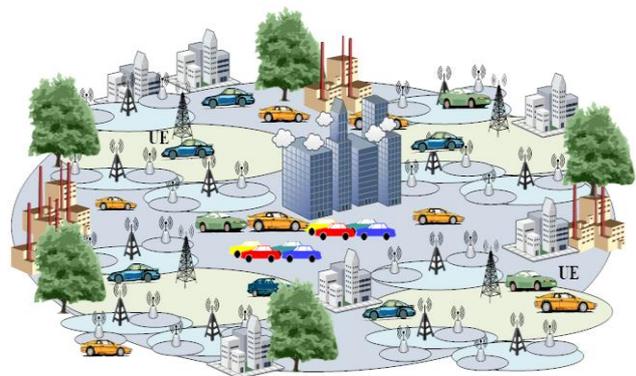

*Fig.1 An exemplary illustration of a vehicular network based on the cellular infrastructure.*

---

\* Corresponding Author: Ilsun You; E-mail: ilsunu@gmail.com

The initial reports on 5G utilize several security functions for authenticating the devices involved in the transmissions [2] [7]. Solutions for authentications, like 5G-Authentication and Key Agreement (AKA) protocol and Extensible Authentication Protocol (EAP)-AKA prime, are discussed as a part of security [2]. Regardless of such discussions, the initial reports ignore the overheads associated with the periodic updates of the secret keys, the range, and mobility of vehicles, as well as the sustainability of V2X against a known set of cyber threats [8][9].

It is a fact that a large number of key updates at regular intervals undoubtedly enhance the security of a network. But such a procedure causes an unintentional burden on the entities leading to high operational cost and computational complexity [6] [10]. Considering all the above discussions and issues, it is desired to further clarify the impact of the key updates and understand its impact on the performance of the network. Moreover, utilizing the existing security functions of 5G and extending them to the edge for V2X security are additional aspects to be resolved. In addition, deriving tradeoffs between the performance and security while grasping the limits of sustainability is a problem to solve, which is not discussed by any of the associated works on 5G-V2X.

Following the above arguments, this article helps to understand the principles of security management for 5G-V2X. It also considers the sub-dividing of the 5G security functions for attaining a high rate of sustainability at a smaller number of key-updates. Furthermore, the entire solution is discussed as a part of a dual security management framework, which uses long-range and short-range based operations and involves sub-divided functions for TMs and hubs. Several theoretical aspects and analytical evaluations facilitate the understandings on the fail-safe points while managing the performance-security tradeoffs. In general, bringing security functions near to the users in 5G-V2X can be theoretically evaluated from this study and feasibilities of key-operations and its sustainability can be observed from the proposed dual security management framework.

The remaining part of this article is organized as follows: Section 2 presents the problem statement and highlights our contribution. Section 3 discusses the related works. Section 4 discusses the deployment model with the system model in Section 5. The proposed approach is presented in Section 6. Section 7 covers numerical evaluations. Finally, Section 8 concludes the article.

## 2 Problem Statement and Our Contribution

Managing security for backhaul-aware 5G-V2X depends on the architectural deployment as well as 5G function-mappings to the underlaid network. Ideally, this is tedious to attain because of the added constraints on operations laid by the devices and the key management. These issues further become a major concern when the key management must be done for edge-initiated devices. Such a focus on bringing security-aspects near to users raises concerns of trustworthy security facilitations, especially when the involved entities are vehicles.

With varying mobility and high dynamics in V2V, V2I and V2P, management of security need an efficient solution that can protect the network from different types of cyber-attacks caused due to non-update of keys and non-compliance of the network for efficient key management. Additionally, the identification of fail-safe points, up to which the network does not require any reconfigurations, is a major challenge.

In this paper, the key exchange passes amongst the vehicles and the 5G-security functions are considered. These passes are then studied for multiple key updates to protect excessive iterations for better performance. Credential management, distributed key-hierarchy principles and edge-initiated security are further considered as a part of the problem.

It should be considered with extreme importance that for the networks including a large set of applications, updating keys after a certain interval becomes a crucial part. Evidently, the network with many key-updates is much secured but with a huge downgrading impact on the performance. Thus, it is required to balance this tradeoff while attaining an efficient backhaul-aware 5G-V2X. Considering the problems defined in the earlier part, the key solutions of this article can be summarized as follows:

- At first, the network is modeled to fit into the requirements of sustainability, which governs the entire system and help to build a backhaul-aware 5G-V2X by re-modeling Impact-Population-Affluence-Technology (IPAT) formulations [11]. This model allows consideration of the independent factors as well as offers flexibility in modeling systems with a simplistic approach as there is not a concurrent model to consider the effects of the vehicular system on the pass involved in the authentications.
- A dual security management framework is presented that considers security through long-range and short-range authentications between the entities.
- Multiple fail-safe points are identified as a part of the optimization solution until which, the network can be operated without changing the derived keys.
- The key hierarchy principle is remodeled for managing the keys attained from the 5G security functions.
- Numerical simulations are conducted to present the tradeoffs as well as understand the efficacy of the proposed system.

## 3 Related Works

The emerging vehicular technologies leverage various solutions emphasized on communication efficiency, mobility and low latency [12]. The 5G era is the adoption of such an advantage that facilitates the service with better connectivity and high speed. The adoption of such technologies is still affected by several issues related to security, privacy, key management, and performance. Due to an increase in the number of security threats, its study and vulnerabilities in the V2X communication received lots of attention from the industrial researchers. The security vulnerabilities can lead to an attack on confidentiality, integrity, authenticity, and availability of network in vehicular communication. Moreover, it affects performance due to improper credential management for vehicles. Several solutions have been proposed by different researchers across the globe aiming at the security and performance of V2X, which is primarily backed by the cellular setups.

Sun et al. [14] proposed a game-theoretic utility model for QoS and security strength in vehicular ad-hoc networks. Their utility model has secured the information capacity based on the security players. Rigazzi et al. [15] proposed a trustworthy communication framework based on Bloom filters to compress the Certificate Revocation Lists (CRLs). It is identified that the significant overhead reduction with low complexity is another challenge for the adoption of dynamic security models in vehicular communications.

Lai et al. [16] focused on the software-defined based secure group communications in the V2X. The group communication enhances road safety and provides high-performance data transmission service for vehicular networks. The authors further discussed the security issues of 5G-VANETs. They also presented a secure and efficient group mobility management framework for the vehicular environment. Such considerations need to focus on handover signaling overheads and latency.

Bian et al. [17] reviewed the security threats against the crowd sensing mechanism in V2X and highlighted the platoon disruption and perception data falsification attacks. Furthermore, Bian et al. [19] investigated the security vulnerabilities in use cases of V2X communications. Ferreira et al. [22] also reviewed wireless vehicular communications based on security properties. Whitefield et al. [18] focused on the privacy preservation of the vehicles. The authors presented Direct Anonymous Attestation (DAA) pseudonym framework and suggested a significant amendment of privacy constraints in the new releasing security documents for V2X.

Fraiji et al. [20] emphasized the security issues of the Internet of Electric Vehicles. The authors discussed security vulnerabilities, replay attacks, false data injection, and jamming attacks. Ahmed and Lee [21] focused on the secure resource allocation in the vehicular networks. Naderpour et al. [23] highlighted the privacy issues in the V2X and suggested an auditable De-anonymization scheme. This scheme is based on the security credential management system and private audit keys. Despite these efforts, the emerging V2X network demands considerable security solutions without affecting the performance of the network. A state-of-the-art comparison, as shown in Table 1, is provided to understand the reachability of the existing solutions.

*Table 1. A state-of-the-art comparison of existing solutions for secure vehicular communications.*

| Approach | Author | Ideology | Parameter/ Concept Used | Security |
|---|---|---|---|---|
| Game-theoretic model for QoS and security strength | Sun et al. [14] | Nash equilibrium | Effect of the frame length, Nakagami fading parameter, total node density, density of eavesdropper nodes | Yes |
| Trustworthy communications framework | Rigazzi et al. [15] | Certificate Revocation Lists (CRL) compression through Bloom filters | Compression gain, Optimal number of hash functions, CRL delivery ratio | Yes |
| Secure group communication | Lai et al. [16] | Software defined technology | Signaling cost, handover latency | Yes |
| Review the security threats against the crowd sensing mechanism in V2X networks | Bian et al. [17] | Platoon disruption attack, perception data falsification attack | Trajectory calculation | Yes |
| Direct Anonymous Attestation (DAA) pseudonym framework | Whitefield et al. [18] | DAA protocol | User-controlled unlinkability and anonymity, Assurance of revocation | Yes |
| Investigate security of V2X communication | Bian et al. [19] | Security threats | Trajectory calculation | Yes |
| Security issues of the Internet of Electric Vehicles | Fraiji et al. [20] | Vulnerability analysis | Replay attack, false data injection, jamming attack | Yes |
| Secure resource allocation scheme | Ahmed and Lee [21] | 3GPP based Random Access with Status Feedback (RASFB) | Success Rate, Resource Utilization, packet reception ratio | Yes |
| Review wireless vehicular communication | Ferreira et al. [22] | Security properties | Incident detection and Recovery | Yes |
| Auditable De-anonymization | Naderpour et al. [23] | Security Credential Management System (SCMS), private audit key | _Tagging and log-chains with lawful identity resolution | Yes |
| Beamforming design for secure V2X | Li et al. [24] | Non-convex optimization problem | harvested power, channel estimation error | Yes |

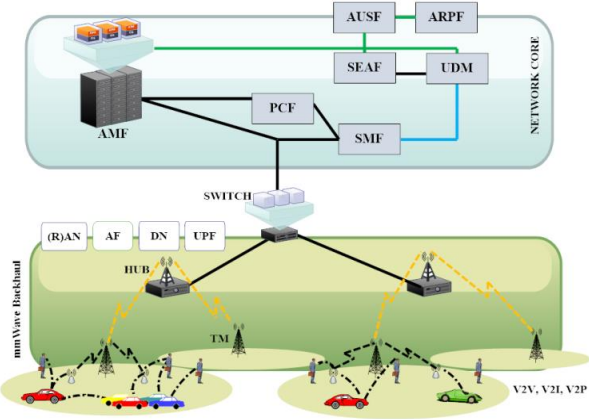

*Fig.2 An exemplary illustration of 5G-V2X with mmWave Backhaul between the Hub and the terminal based on the 5G security and general network functions initially presented in [13] (AMF: Access and Mobility Management Function, PCF: Policy Control Function, AUSF: Authentication Server Function, ARPF: Authentication Credential Repository and Processing Function, SEAF: Security Anchor Function, SMF: Session Management Function, UDM: Unified Data Management, AF: Application Function, (R)-AN: (Radio) Access Network, DN: Data Network, UPF: User Plane Function, TM: Terminal, UE: User Equipment (Vehicle)).*

## 4 Deployment Model

The network is derived using the existing 5G security functions, which helps to map the edge devices to the core. The backhaul is dominated by the mmWave technology, which helps to maintain a high speed of accessibility for the User Equipment (UE), as shown in Fig.2. The UEs include the vehicles which are the observers in the V2V, V2I and V2P formations. The core comprises the Access and Mobility Management Function (AMF), which operates together with the Session Management Function (SMF) and Policy Control Function (PCF), and the Security Anchor Function (SEAF) is placed in their periphery which is located deep in the network without any knowledge to the edge as well as vehicles as stated in [2]. The present draft is unable to provide any proven support for deploying edge-initiated security as well as the distributed authentication for V2X, which is attained through the proposed solution.

The proposed model presents a new disintegrated structure to make the 5G-drafted version suitable for V2X based on key-divisibility. In the considered model, the SEAF keys and hierarchical authentication are driven by the Authentication Server Function (AUSF) and the Authentication Credential Repository and Processing Function (ARPF) [2]. To make the 5G-V2X supportable for backhaul and provide additional situational awareness, the mmWave technology is considered for supporting the Unified Data Management (UDM), Application Function (AF), (Radio) Access Network ((R)-AN), and User Plane Function (UPF).

## 5 System Model

As stated in the problem statement, the security management for V2X is tedious as it involves multiple authentications. This issue becomes further complex when we are to deploy the security functions at the edge of the network. This procedure involves complex modeling, which might affect the cost of the network.

The proposed approach operates by considering the security directly proportional to the number of key-updates performed in the V2X between the TMs and the Core. With the number of updates growing extensively, the network suffers from several induced overheads, which slow down the operations. Moreover, it also increases the burden of credential management. To resolve this, at first, a timing problem is formulated around the operational period of the network.

For this, let $t$ be the time taken by an adversary to launch an attack, out of which let $t'$ be the minimum time for which the keys should not be updated. Considering this, the key-utilization time, $t_u$, is identified, such that $t_u < t'$ $(<=t)$, for which the currently allocated keys are considered safe from the known attacks. Moreover, intermediate-key updates, $U_k$, should be minimized while maximizing the overall sustainability, $S_N$, of the network, where, $S_N$ is derived as a function of induced overheads by altering the IPAT formulations [11], such that

$$S_N = \frac{nU_k}{D.P.Q}, \qquad (1)$$

Here,

$$D (\leq N) = \int_{r1}^{r2} C(x,y) dx \qquad (2)$$

is the number of cars in the periphery (r) of a particular TM for a given density function $C(x,y)$, $N$ is the number of vehicles and $E$ is the total number of entities operating in the network, $P$ is the probability of loss in connectivity, which is calculated as $1 - \frac{E'}{E}$; where $E'$ is the number of transmitting/connected entities and $E' \leq E$. $Q$ is the number of passes between the entities. It is calculated as the number of messages shared between the involved entities and is dependent on the protocol used for sharing final data between the devices. Under strict evaluations, it may also add up the signaling messages as well as the messages used for re-authentication. For generic evaluations, the number of messages used by a protocol from initiating to full control is considered as a value of $Q$. Here, $n$ is the sustainability balancing constant depicting the inverse of the number of hops between the vehicles and the entity dealing with the request generated for a 5G security function (key or service requests). These parameters can be used to resolve the above-discussed security issues by formulating a below-given optimization problem:

$$\max(S_N) \, \forall E, \forall N, \qquad (3)$$

s.t.
$$\max(t_u), \forall N,$$

$$\underbrace{\min(U_k)}_{\text{in tradeoff with } S_N}, \text{ and } U_k \geq U'_N,$$

$$0 < D \leq N,$$

$$0 < \frac{n^{-1}(n^{-1}-1)}{2} \leq \frac{E(E-1)}{2}, n^{-1} \neq E,$$

$$\min(t_u - t'). \qquad (4)$$

In (4), $U'_N$ is the mandatory key-updates below which the network security is difficult to evaluate and additional tasks such as mobility management, re-authentication cannot be performed efficiently. Furthermore, the tradeoff between the performances in terms of sustainability can be balanced for attaining the security of operations by considering the range of authentication. This means that the distance of the key update (R) should be managed to prolong the connectivity of the network without affecting the probability of loss in connectivity (P). Overall, $P \propto R$ i.e., with an increase in the operational range, the loss in connectivity may increase, which should also be managed by the approach. Considering these, an optimal solution should aim at the selection of suitable points when key updates should be minimized and changed to maintain sustainability. For this, (1) can be remodeled to $\max(S_N^{(R)})$, which is given by

$$S_N^{(R)} = \frac{nU_k}{D.R.Q}, \qquad (5)$$

with the optimization principle following the same conditions in (3) ~ (4) additionally at $\min(P)$; and $\max(1 - p_x)$, which refers to the availability of credentials.

# 6 Proposed Approach

The proposed approach forms a dual management framework, which relies on two modes for establishing security in 5G-V2X. These modes are driven by distance as well as timing for the management of keys and maintaining a secure network and are termed as long-range authentication and short-range authentication. Both these modes facilitate a strong authentication for V2X while ensuring the security of backhaul between the TM and the hub. The proposed approach relies on key exchanges that are performed by utilizing the deployed sub-divided 5G security functions.

The backhaul links are facilitated by the long-range operations whereas the edge-enabled formations are supported by the short-range authentication. Both these modes further allow the timely management of keys and help to ensure the optimal solution for identification of the fail-safe points based on certain derivations presented in the lateral part of this section.

The initial mode is based on the derivations of keys from the core, which is a part of long-range operations and then sub-division of initially obtained keys to support the procedures for short-range authentication.

The long-range authentication uses mmWave communications, which must be facilitated by existing encryption algorithms. The long-range operations update the keys once the devices are deployed or reconfigured. Usually, existing solutions can support this operation; however, based on the speed of operations of mmWave, the associated operations of encryption and decryption should be performed in a timely manner to avoid any undue overheads. At present, this article deals with the identification of slots, i.e., when the associated keys should be updated and when the network should be reconfigured. The formation of the long-range authentication process is assumed and is beyond the scope of this article.

The major issue prevails at the edge-enabled vehicles, which under high mobility, cause an additional overhead of key-regenerations as well as credential management. To further quantify this issue, a factorized module $(G_f)$ is considered in the short-range model of V2X, which evaluates the speed (S), location (L), last updates for keys $(U_T)$, shared sessions $(A_S)$, refreshing rate of keys $(F_R)$, total keys $(T_K)$, zone traversals $(Z_T)$, and associatively $(V_A)$ of a vehicle to generate new keys or continue with the existing keys while maintaining the conditions stated in (3)~(5). According to which, $G_f = f(S, L, U_T, A_S, F_R, T_K, Z_T, V_A)$, where value of the function can be modeled with the help of priority constants and each value can help to determine the factor(s) which should be given importance while deciding the slots for key updates as well as for setting fail-safe points. In general, $G_f$ can be set as:

$$G_f = \frac{1}{\sum_{i=1}^{w} \theta_i} \sum_{i=1}^{w} g_i \theta_i. \qquad (6)$$

Here, $w (\leq 3)$ is the number of subdivided components (g) for the given $G_f$, which are given by:

$$g_1 = \delta_1 f(S, L), \qquad (7)$$

$$g_2 = \delta_2 f(U_T, A_S, F_R, T_K), \qquad (8)$$

$$g_3 = \delta_3 f(Z_T, V_A), \qquad (9)$$

where $\delta_1$, $\delta_2$ and $\delta_3$ are the distribution constants for the associated functions in (7), (8) and (9), respectively with their sum in range 0 and 1; and $\theta$ is the associated priority constant and also has its value between 0 and 1 depending on the dependencies. At present, the proposed approach does not account for their in-depth evaluations and assumptions are considered to check for other major aspects of the number of passes and fail-safe points. These functions are dependent on the configuration of the network and the associated distribution changes based on the configurations.

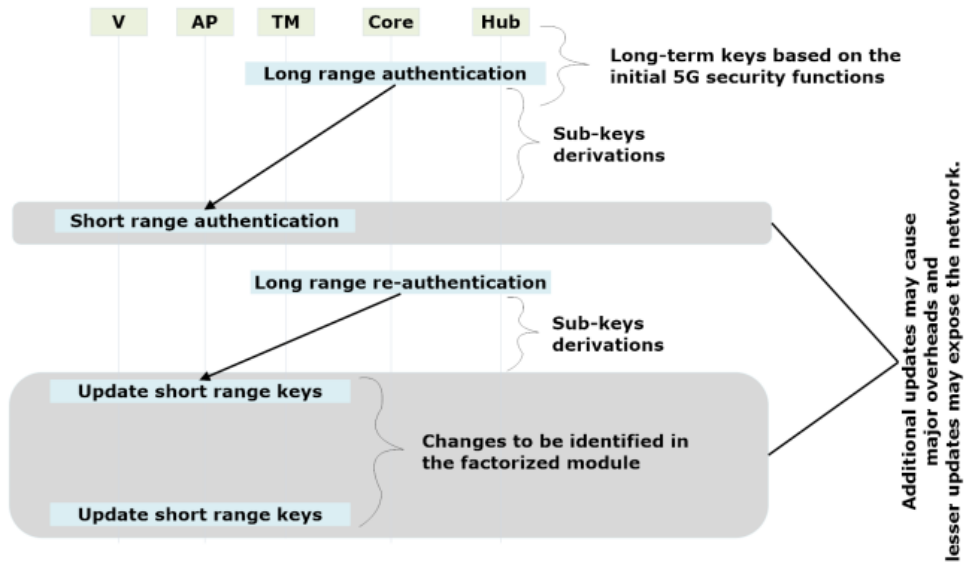

*Fig.3 An overview of the security considerations with long and short-range authentications (V: Vehicle (UE), AP: Access Point, TM: Terminal).*

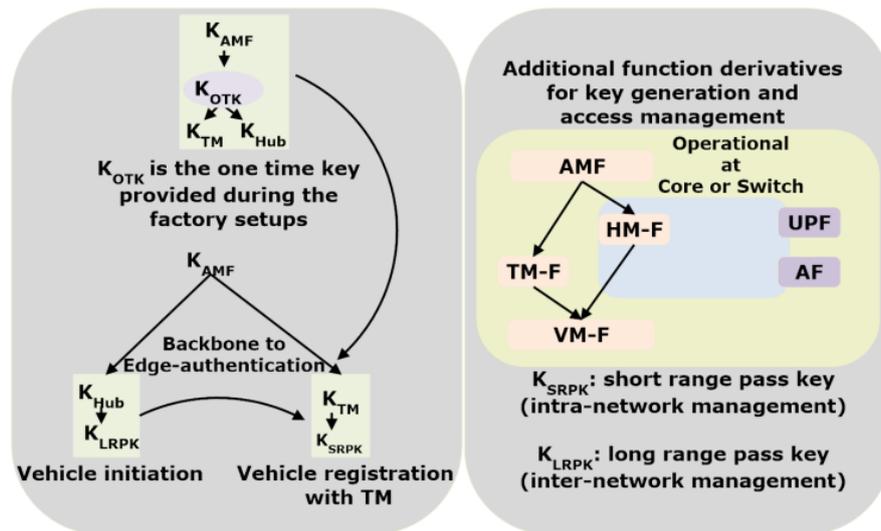

*Fig.4 An overview of the key generations, long and short-range authentication and additional functions derived for managing sub-keys.*

### 6.1 Dual Security Management and Key Derivations

An overview of the proposed approach along with key dependencies is presented in Fig.3 and Fig.4. The long-range is operable between the TMs and the Hubs through the core. The long-range authentications are carried out using traditional 5G security functions, which uses SEAF and SMF coordination to follow security implementations. The original functions are kept exactly similar for long-range authentications, and re-authentication, when the network undergoes substantial changes, is done through these passes. The long-range security functions follow the sub-dividing of original functions each time some major changes are made to the network setup. Moreover, it is flexible to decide if the changes made to the system are subject to revivification by the core.

In certain cases, long-range authentications can only be used in the initial phases and then the entire security can be formulated based on certain long-term shared keys, which allows security throughout the time an entity is active in the network. However, as required, the key aspect is to handle the security of vehicles near users with lesser overheads. The proposed approach relies on the short-range authentication which uses key-controlling feature and identification of fail-safe points to make sure that the network is aloof from the certain known cyber threats. Moreover, it helps to bring 5G security functions near the users at the edge by sub-dividing the functionalities, as shown in Fig.4. The description of sub-divided functions is as follows:

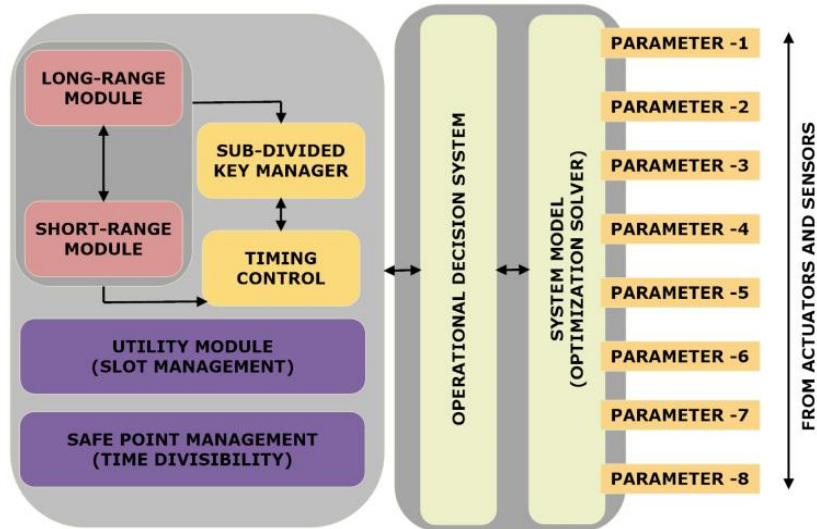

*Fig.5 Proposed dual security management framework with optimal slot identification and fail-safe markings.*

- The AMF function is further divided into two functions- the terminal management function (TM-F) and the Hub management function (HM-F) that derive their keys from $K_{OTK}$. Here, $K_{OTK}$ is the one-time key, which is derived from $K_{AMF}$, where $K_{AMF}$ is obtained similarly to the regular 5G security modules. $K_{OTP}$ is provided during the factory setup phases and is used once the entity is deployed in the network. It plays a key role in the long-range authentications. Any protocol aiming at long-range authentications can use this $K_{OTP}$ as a one-time shared key for observing a secure route between the entities in the initial phase. This $K_{OTP}$ can also be set to regeneration during reconfigurations or new keys can be derived to ensure the security of entities when the controlling hubs or switches are reallocated/reinitiated/rebooted.
- $K_{OTP}$ is used to drive the terminal management key ($K_{TM}$) and the hub management key ($K_{Hub}$). These keys are used for establishing the backhaul to edge connections and maintain the session based on the linkage with the $K_{AMF}$.
- Further, $K_{Hub}$ and $K_{TM}$ are used to derive the operational keys, namely, $K_{LRPK}$ and $K_{SRPK}$, i.e., long-range passkey and short-range passkey, respectively. The derived keys are used for initial long-range authentications and regular short-range authentications. The sub-derived keys authenticate vehicles' fronthaul to the edge and the backhaul between the TM and the hub.
- The sub-derivative functions help to support the intra- and inter-mode of secure communications in V2X based on the nature of the passkey and its controlling entity as expressed in Fig.4. All the generic operations of these keys are driven by their respective core operations available from the 5G security functions.
- The HM-F and the TM-F operate at the edge controlled by their parent function in the core and are considered aligned in the same layer as that of UPF and AF. In this way, the security controls are shifted near to the edge without compromising with the requirements of settings laid in the original 3GPP draft. Moreover, in the worst cases of zero-day attacks only the edge-entities are exploited, which can be reconfigured as well as replaced at fewer expenses compared with the entire remodeling of the core.

In the proposed model, it can evidently be observed that with short-range authentications, the number of key operations becomes extensively large which is a core requirement of all the approaches aiming at edge-initiated security. In such a scenario, the overheads are bound to happen, and it can be controlled by the defined problem statement, which makes the proposed solution practically feasible as well as important. Additionally, it is observable that the proposed approach operates in two modes of authentication, which makes it a dual security management model whose details on formations can be observed in the next subsection.

### 6.2 Overall Framework and Decision Module

The proposed approach can be modeled into a combined framework as shown in Fig.5. The proposed framework includes several sub-modules, which finally uses a decision module to control the instances when the keys should be updated while maintaining a check on the sustainability of the network. The overall framework includes the following modules:

- Long-range module: This module is responsible for managing the connectivity between the core and the device which forms the backhaul in the network. This module also controls the TMs and the hubs and additionally supports the coordination between the edge-side and the core-side 5G security functions. The long-range module interacts with the sub-divided keys, which are responsible for the security coordination of the vehicle with the edge network. This module is operable on the vehicle as well as network infrastructures.

- Short-range module: This module operates in the lower periphery of the long-range module and is maintained on-board of all the vehicles. In certain cases, where cost is not an issue, this module can be used on infrastructures. However, in such a situation, the cost of security and associated algorithms' complexity may increase by many folds. Thus, in the usual scenario, the short-range module is subject to implementation only on the vehicles' interface.
- Sub-divided key manager: This part of the framework manages the number of keys associated between the edge and the vehicles. This module maintains the hierarchy of keys and helps in derivation as well as updating whenever the requests are generated from the timing control.
- Timing control: This part operates together with the short-range module and the sub-divided module and is invoked as per the decisions available on the sustainability of the network. The timing control helps to maintain a limit on the number of updates a system performs while operating vehicles as end users in 5G-V2X.
- Utility module: This module manages the slots which are observed from the timing control. The logs are observed in this part. This also helps to track the instances when the network faced vulnerabilities and was under threat by several cyber-attacks.
- Safe point management: This module is the actual carrier of the network when the checks are observed based on the values of sustainability. This module helps to limit the number of key operations by operating in coordination with the timing control and the utility model.
- Operational decision system: This module takes the final decision by evaluating the long-range and short-range authentications. It helps to decide whether the updates are required in the keys or the network can be operated at the same configurations.
- Optimization solver: This part is responsible for giving inputs to the system by using the factorized modules which is a combination of several parameters as used in the derivation of $G_f$. This module is dynamic and can be extended based on the available set of parameters with their values from the underlying sensors and actuators.

The entire procedures of the dual security management framework can further be observed from the following Lemmas:

**Lemma-1:** $S_N$ is divergent if the initial information on the key generation is unavailable. However, in the case of known timestamps, the network sustainability can be modeled over the available key updates and vehicle movement, such that for instances $t_1$ and $t_2$:

$$S_N = \frac{\alpha^2}{2\beta N \left(1 - \frac{n^{-1}}{E}\right)^N Q} \left(Ei\left(\frac{\beta - \alpha}{t_1}\right) - Ei\left(\frac{\beta - \alpha}{t_2}\right)\right),$$
(10)

at $t_2 - t_1 > 0$, $E - n^{-1} > 0$, $\beta - \alpha > 0$, and $Ei$ is the exponential integral [13].

**Proof:** Considering (1) to be following Poisson distribution for vehicles approaching at a rate of $\beta$ vehicles per unit time and operating with $\alpha$ number of key-updates per unit time, $U_k$ can be given by $e^{-\frac{\alpha}{t}}\frac{\left(\frac{\alpha}{t}\right)^X}{X!}$, (X=2), as only two keys are used for authentication (long range and short range keys) and $D$ can be written as $e^{-\frac{\beta}{t}}\frac{\left(\frac{\beta}{t}\right)^{X'}}{X'!}$, (X'=1), as each vehicle maintains its connectivity with one source from the network. Now, considering the involved entities in the network for authenticating a vehicle, and using the model in [12], the probability of no connectivity (P) can be given as $\left(1 - \frac{n^{-1}}{E}\right)^N$. The number of passes remains at the discretion of the used protocol and is constant for this evaluation. By using these values in (1) and under fixed interval, the observed equation can be written as:

$$S_N = \frac{1}{N\left(1-\frac{n^{-1}}{E}\right)^N Q} \int_{t_1}^{t_2} \frac{e^{-\frac{\alpha}{t}\frac{\left(\frac{\alpha}{t}\right)^2}{2!}}}{e^{-\frac{\beta}{t}\frac{\left(\frac{\beta}{t}\right)^1}{1!}}} dt .$$
(11)

On solving, at $t_2-t_1>0$, $E-n^{-1}>0$, $\beta - \alpha>0$, the observation is

$$S_N = \frac{\alpha^2}{2\beta N \left(1-\frac{n^{-1}}{E}\right)^N Q} \left(Ei\left(\frac{\beta-\alpha}{t_1}\right) - Ei\left(\frac{\beta-\alpha}{t_2}\right)\right),$$
(12)

which is the desired output [13].

**Lemma-2:** For extreme large rates of key exchanges and vehicle dynamics, $S_N$ is convergent to a linear function under asymptotic observations. This can be further used to identify fail-safe points with high accuracy up to which the network can be operated without much overheads and security breaches [13].

**Proof:** In continuation from Lemma-1, the result in (12) can be asymptotically analyzed for evaluating the behavior of the curve determining the sustainability of the V2X under given constraints. According to which, if $\alpha$ and $\beta$ increases to a larger value, the (12) reduces to a linear function, such that $S_N$ can be evaluated as a function comprising two metrics, i.e. $S_N = f\left(\frac{\alpha}{t}, \frac{\beta}{t}\right)$; and by following strict principles and limits, it can be given as $S_N = \frac{\alpha}{\beta}$. Moreover, in such a case, the behavior is not integrally divergent, and the network can be evaluated with linear timestamping. Now, considering these observations,

the network fail-safe points up to which there is no need to update or refresh the keys can be evaluated as:

$$F_S = \begin{cases} t \text{ at } S_N \geq S_N^{TH}, \text{if } t_1 \neq 0, \text{and known} \\ t \text{ at } M_O \leq M_O^{TH}, \text{otherwise} \end{cases}. \quad (13)$$

Here, $M_O$ is the message overhead evaluated as

$$M_O = \frac{O_S(1-P)}{E \cdot P}, \quad (14)$$

where $O_S$ (signaling overheads) is said to be increasing exponentially. This is due to the reason that the key updates and vehicles follow the Poisson distribution with randomness observable from Lemma-1, based on which, $O_s$ can be modeled as $O_b(1-\frac{\alpha}{t})^t$, where $O_b$ is the overheads for initial authentication. This can be predicted between the instances, as shown in Lemma-1, only at a fixed number of updates ($\alpha' = \frac{\alpha}{t}$), such that

$$O_S = \frac{O_b\left(\frac{n-1}{E}\right)^N}{E\left(1-\frac{n-1}{E}\right)^N} \cdot \frac{\ln(1-\alpha')^{t_2} - \ln(1-\alpha')^{t_1}}{\ln(1-\alpha')}, \quad (15)$$

which is the desired output [13].

**Lemma-3:** The prediction on message overheads allows the network to take pre-hand decisions on the number of updates which would be enough to keep the system operating at the given rate. Such a prediction is based on the range of vehicles mobility during the time given time slots, and it driven by the probabilistic states, such that

$$M_O^{(predicted)} = \frac{O_b\left(\frac{n-1}{E}\right)^N \left(\ln\left(1-\frac{\alpha}{t_2}\right)^{t_2} - \ln\left(1-\frac{\alpha}{t_1}\right)^{t_1}\right)}{\ln\left(1-\frac{\alpha}{t_2}\right) E(e^{-\frac{\alpha}{t_2}} - e^{-\frac{\alpha}{t_1}})} \cdot \frac{\alpha(r_2-r_1)}{\beta N\left(1-\frac{n-1}{E}\right)^{2N}} \left(Ei\left(\frac{\beta-\alpha}{t_1}\right) - Ei\left(\frac{\beta-\alpha}{t_2}\right)\right) \cdot \left(1 - \frac{E_0}{E\gamma'}(e^{-\gamma' t_1} - e^{-\gamma' t_2})\right) \quad (16)$$

at $t_2 - t_1 > 0$, $t_2 > 0$, $t_1 > 0$, $r_2 - r_1 > 0$, $r_2 > 0$ and $n^{-1} \neq E$.

**Proof:** In the given model for (1), Beta function [25] can be used to model the traffic as it allows better prediction on the state of the network and allows accurate identification of the likelihood of the existence of a vehicle in the defined range, such that,

$$D = f(C(\vartheta, \mu)) = \int_{r_1}^{r_2} \frac{\Gamma(\vartheta+\mu)}{\Gamma(\vartheta)\Gamma(\mu)} \gamma^{\vartheta-1}(1-\gamma)^{\mu-1} d\gamma \quad (17)$$

where $\vartheta$ and $\mu$ are the shape and scale parameters for the incoming rate $\gamma$ of the vehicles in the range $r_1$ and $r_2$. The system operates with $\vartheta \geq 1$ and $\mu = \frac{E}{\sum_{i=1}^{E}(\log(\frac{1}{1-p_x}))_i}$. Here, $1 - p_x$ is the probability of availability of credentials and it can be operated only for vehicles ($N$) instead of the entire involved entities in the network given by $E$. Note that the Beta function is omitted under special circumstances when only the number of key-updates is controllable and vehicle's actual movement is unpredictable. In the given prediction model, $U_K$ can be predicted as $\int_{t_1}^{t_2} e^{-\frac{\alpha}{t}} \frac{(\frac{\alpha}{t})^2}{2!} dt$, $S_N$ can be predicted from (10) and $O_s$ can be observed from (15). Here, $P (= 1 - P_c)$ can be predicted based on the failure rate of the entities in the network such that

$$P_c = \frac{E - (E_0 e^{-\gamma' t})}{E}, \quad (18)$$

where $\gamma'$ is the outgoing rate of the vehicles in the defined range during an instance, and $E_0$ is the initial number of vehicles with $E_0 \leq E$. Now, by using outcomes of Lemma-1 and Lemma-2, the message overheads can be predicted by using (1) and (14) as:

$$M_O^{(predicted)} = \frac{Q}{U_K^{Predicted}} \left(S_N^{Predicted} \cdot O_S \cdot D^{Predicted} \cdot (1 - P^{Predicted})\right). \quad (19)$$

This on solving at unit shape parameters gives

$$M_O^{(predicted)} = \frac{O_b\left(\frac{n-1}{E}\right)^N \left(\ln\left(1-\frac{\alpha}{t_2}\right)^{t_2} - \ln\left(1-\frac{\alpha}{t_1}\right)^{t_1}\right)}{\ln\left(1-\frac{\alpha}{t_2}\right) E(e^{-\frac{\alpha}{t_2}} - e^{-\frac{\alpha}{t_1}})} \cdot \frac{\alpha(r_2-r_1)}{\beta N\left(1-\frac{n-1}{E}\right)^{2N}} \left(Ei\left(\frac{\beta-\alpha}{t_1}\right) - Ei\left(\frac{\beta-\alpha}{t_2}\right)\right) \cdot \left(1 - \frac{E_0}{E\gamma}(e^{-\gamma' t_1} - e^{-\gamma' t_2})\right), \quad (20)$$

which is the desired output at $t_2 - t_1 > 0$, $t_2 > 0$, $t_1 > 0$, $r_2 - r_1 > 0$, $r_2 > 0$ and $n^{-1} \neq E$.

**Lemma-4:** If $S_N$ and $M_O$ are to be calculated based on the available information, the fail-safe points can be accommodated as in (13); However with the variation in $S_N$, the fail-safe points are modeled as per the given distribution, and can be predicted based on current distribution, i.e., if Beta function is used as shown in Lemma-3, the likelihood of occurrence of fail-safe points ($\tau$) between $t_1$ and $t_2$ can be given as:

$$\tau = \arg\max \frac{\Gamma(1+\mu)}{T \Gamma(\mu)} \cdot \frac{\frac{(d_2-1)^2}{(1-d_2)^\mu} - \frac{(d_1-1)^2}{(1-d_1)^\mu}}{\mu-2}, \quad (21)$$

assuming that $\mu - 2 > 0$, $\vartheta = 1$, $d_2 - d_1 > 0$, $1 - d_2 > 0$ and $1 - d_1 > 0$.

**Proof:** Let $1 - \omega_x$ be the probability that the available value of $S_N$ is following the given constraints of the system to keep on without changing or updating the current keys; $\varphi$ is the observable checkpoints in the system referring to the probabilistic time for updating keys. Now, considering similar distribution as that of vehicles in (17), the likelihood of fail-safe points ($\tau$) can be accommodated as

$$\arg\max \frac{1}{T} \int_{d1}^{d2} \frac{\Gamma(1+\mu)}{\Gamma(\mu)} (1-\varphi)^{\mu-1} d\varphi, \quad (22)$$

at unit shape parameter such that $\mu = \frac{mean(S_N)}{\sum_{i=1}^{t_x}(log(\frac{1}{1-\omega_x}))_i}$, where $t_x$ ($\leq t_u$) is the known time slots for which average $S_N$ can be observed. This on solving gives,

$$\tau = \arg\max \frac{\Gamma(1+\mu)}{\mathrm{T}\,\Gamma(\mu)} \cdot \frac{\frac{(d_2-1)^2}{(1-d_2)^\mu} - \frac{(d_1-1)^2}{(1-d_1)^\mu}}{\mu-2}, \quad (23)$$

which is the desired output at $\mu - 2 > 0$, $\vartheta = 1$, $d_2 - d_1 > 0$, $1 - d_2 > 0$ and $1 - d_1 > 0$. Now, assuming that the values of $S_N$ are separable for each of the vehicle irrespective of the number of vehicles operating in the same range, the likelihood of occurrence of a fail-safe point can be simply observed equivalent to $\omega_x$ by following the principles of likelihood defined in [26].

**Lemma-5:** In the given model, the scale parameter, $\mu$, decides the operations of updating keys. Furthermore, it helps to fix the thresholds for deciding $U'_N$ when a vehicle is active for a long duration in the same zone. Thus, it becomes important to understand its asymptotic nature, which by using asymptotic equality [27] can be given as:

$$f(\mu) = \int_{c1}^{c2} \frac{e^{-\frac{\gamma'}{T}}}{log(\frac{1}{1-\gamma'})} d\gamma', \quad (24)$$

at with $c_1 \geq 0$, $c_2 - c_1 > 0$, $c_1 < 1$ and $c_2 < 1$. From which it can be observed that the true value of $U'_N$ can never be predicted for a particular vehicle and it has to be assumed or alternatively derived using $g_l$, which is the function of location and speed.

**Proof:** By considering $\mu = \frac{E}{\sum_{i=1}^{E}(log(\frac{1}{1-p_x}))_i}$ approximately equivalent to $\frac{mean(S_N)}{\sum_{i=1}^{t_x}(log(\frac{1}{1-\omega_x}))_i}$, the scale parameter can be subjected to modeling based only on the outgoing-rate of vehicles instead of incoming rate when the minimum number of key updates is to be calculated. According to which, the above-discussed conditions can be written as:

$$\mu = \frac{E}{\sum_{i=1}^{E}(log(\frac{1}{1-\gamma'}))_i}, \quad (25)$$

which on following the discussions in Lemma-3 can be reduced to:

$$\mu = \frac{1}{log(\frac{1}{1-\gamma'})}, \quad (E = E_0 e^{-\gamma' t}). \quad (26)$$

The above condition can be converted to asymptotic equality using [27], such that $f(\mu)$, which is the asymptote of scale parameter, is given as:

$$f(\mu) = \int_{c1}^{c2} \frac{e^{-\frac{\gamma'}{T}}}{log(\frac{1}{1-\gamma'})} d\gamma', \quad (27)$$

which is the desired observation with $c_1 \geq 0$, $c_2 - c_1 > 0$, $c_1 < 1$ and $c_2 < 1$. However, under the given limits, the integral is divergent, which can be converged by following linear modeling for the scale parameter by using speed and location as the input metrics. Furthermore, the expression in (27) can be expanded for exponential asymptote, according to which, the asymptote behavior of $\mu$ can be modeled as $T^{t_x}$.

The above observations are required to find the exact slots for updating the keys based on the available values of sustainability as well as message overheads. The thresholds are network and protocol dependent. These are identified based on the information of the time required to launch an attack on the deployed protocols. Especially, the general thresholds are motivated by the algorithms operating for the authentication of V2X or TM to the hub.

*Table 2. Numerical simulation settings*

| Metrics | Values |
|---|---|
| $\beta$ | 2-10 (step 2) (A1-A5) |
| $\alpha$ | $\beta/2$ |
| $N$ | 10 |
| $E$ | 10-50 |
| $n^{-1}$ | 5 |
| $Q$ | 1 to 5 |
| $t_1$ | 5 s |
| $t_2$ | 105 s |
| $T$ | 110 s |
| $\gamma'$ | 0.1 to 0.9 |
| $r_1$ | 100 m |
| $r_2$ | 500 m |
| $c_1$ | 0.1 |
| $c_2$ | 0.9 |
| $\vartheta$ | 1 |
| $t_x$ | Step by 5 s |
| $1 - p_x$ | 0.1 to 0.9 |
| $1 - \omega_x$ | 0.1 to 0.9 |
| $d_1$ | 0.1 |
| $d_2$ | 0.9 |

## 7 Numerical Evaluations

The security of any network cannot be quantified but it can be predicted based on the network performance by assuming certain aspects of the system modeling. Based on this argument, the timing to update keys and managing security based on the incoming/outgoing rate of vehicles is quantifiable as expressed through Lemma-1 to Lemma-5. To further understand the impact of metrics on the sustainability and the location of fail-safe points, a numerical case study is conducted by using values defined in Table 2.

The results for sustainability from Lemma-1, as shown in Fig. 6, suggest that the proposed approach allows effective tracing of the network activity and manage its security by keeping a check on the unnecessary key updates.

These results suggest that significant overheads can be reduced from the network while keeping the security at the edge of the network by utilizing sub-divided functions for TMs and hubs. The protocol dependent $Q$ and $U_k$ used for authentication pose a significant impact on the performance as well as the security of the network.

Quantitatively, the value for sustainability varies between 53.1% and 84.9% compared to initial observed output at constant arrival and update rates for vehicles and keys, respectively.

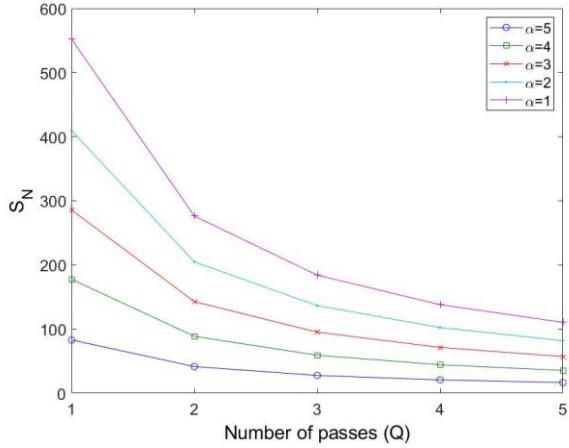

*Fig. 6 Network sustainability w.r.t. the number of protocol-passes at different rates for incoming vehicles and the number of key updates.*

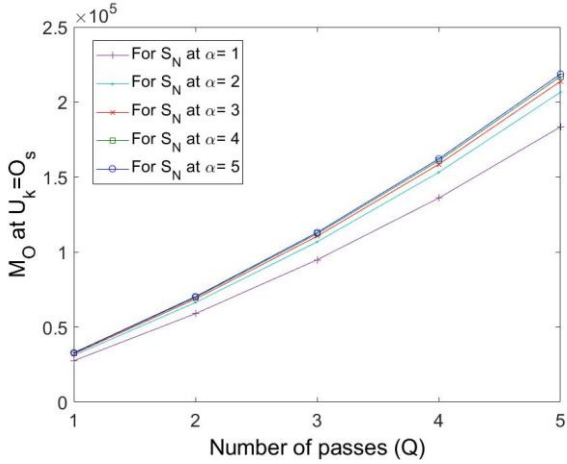

*Fig. 7 Message Overheads ($M_O$) vs. the number of protocol-passes (Q).*

The results for message overheads are provided in Fig.7. These results are tracked based on the Lemma-2 by varying $S_N$ for the different range of α. With an increase in the number of passes, it is obvious that the message overheads are bound to increase. However, with the control over the arrival rate of requests, $M_O$ can be controlled to go beyond a limit. The results show a variation between 11.1% and 16.1% for a varying range of sustainability provided in Fig.6. These results help to understand the limits of the fail-safe operations when the proposed dual security management is used for 5G-V2X.

Similarly, the results are observed for the failure of probability by varying the vehicle outgoing rate at different step interval for vehicles, as shown in Fig.8. With vehicles moving at a faster pace, the probability of failure increases as the connections for sustainability decreases and it becomes difficult to manage the network. Such a condition also poses a significant impact on the credential management of the 5G-V2X. The results show that the average value of $P$ varies between 0.83 to 0.99 for steps varying between 5 and 25. In addition, the prediction of the system shows an average observation between 0.39 and 0.98, as shown in Fig. 9. This prediction is in the range of ±11.7%. Furthermore, the scale parameter can be traced for varying values of $\gamma'$ and $\omega_x$, as shown in Fig. 10.

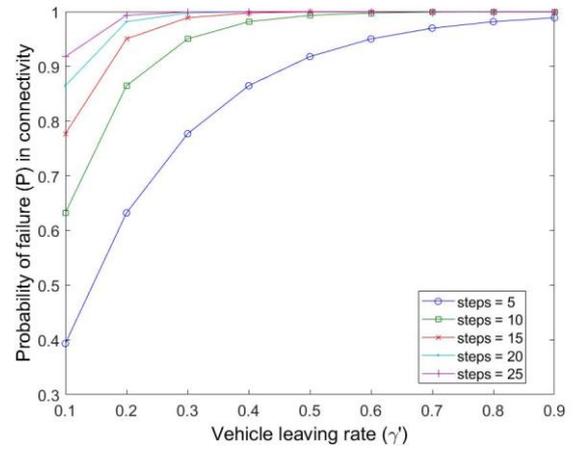

*Fig. 8 Probability of failure (P) in connectivity vs. vehicle leaving (outgoing) rate.*

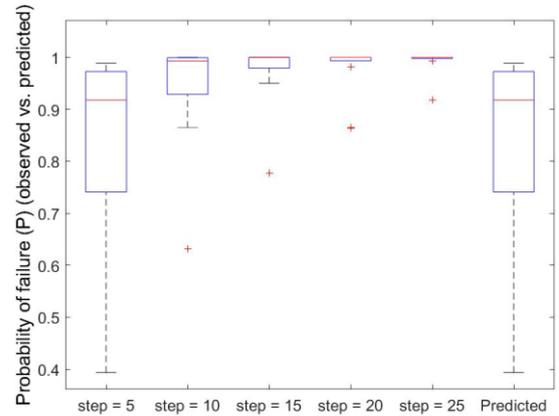

*Fig.9 Probability of failure (P) - prediction vs. observed values at varying timestamps.*

The graph shows the conditional value depicting that the network cannot perform if $\mu \leq 2$. The network at any time, when observing such scale values, undergoes reconfigurations leading to mandatory key updates. The management of the scale parameter can control the performance of the system as it affects the probability of connectivity in the proposed model. In an extension to these results, the scale parameter can be observed for varying values of $Q$, as shown in Fig.11.

Table 3. Fail-safe durations for secure 5G-V2X for the proposed dual security management framework.
(*$d_2$ should be adjusted to accommodate high scaling values in the network; alternatively, the time
steps need to be increased, which is again the subject of optimization and extremely crucial as
discussed in the analytical model.)

| Cases | $\mu$ Q=1 | $\mu$ Q=2 | $\mu$ Q=3 | $\mu$ Q=4 | $\mu$ Q=5 | $\tau$ (Q=1) | $\tau$ (Q=2) | $\tau$ (Q=3) | $\tau$ (Q=4) | $\tau$ (Q=5) |
|---|---|---|---|---|---|---|---|---|---|---|
| Optimistic Scenario (Non-realistic values) | 659.08 | 329.54 | 219.68 | 164.76 | 131.80 | * | * | 4.39E+215 | 5.3E+160 | 5.9E+127 |
| | 311.20 | 155.60 | 103.73 | 77.79 | 62.23 | * | 3.65E+151 | 4.92E+99 | 5.8E+73 | 1.61E+58 |
| | 194.69 | 97.35 | 64.89 | 48.67 | 38.93 | 4.51E+190 | 2.056E+93 | 7.326E+60 | 4.43E+44 | 8.24E+34 |
| | 135.94 | 67.97 | 45.31 | 33.98 | 27.19 | 8.03E+131 | 8.736E+63 | 1.943E+41 | 9.28E+29 | 1.5E+23 |
| Practically feasible Scenario (realistic values) | 100.18 | 50.09 | 33.39 | 25.04 | 20.03 | 1.41E+96 | 1.169E+46 | 2.385E+29 | 1.09E+21 | 1.09E+16 |
| | 75.79 | 37.89 | 25.26 | 18.95 | 15.16 | 5.70E+71 | 7.496E+33 | 1.797E+21 | 8.96E+14 | 1.5E+11 |
| | 57.68 | 28.84 | 19.22 | 14.42 | 11.53 | 4.47E+53 | 6.734E+24 | 1.701E+15 | 2.77E+10 | 37629917 |
| | 43.15 | 21.57 | 14.38 | 10.79 | 8.63 | 1.34E+39 | 3.75E+17 | 2.54E+10 | 6816294 | 50296.62 |
| | 30.16 | 15.08 | 10.05 | 7.54 | 6.03 | 1.40E+26 | 1.257E+11 | 1279251.7 | 4280.43 | 146.0569 |
| | - | - | - | - | 2 | - | - | - | - | 0 |
| | - | - | - | - | 1 | - | - | - | - | 0 |

The results suggest that with a lower number of protocol overheads, the security of the system increases as lesser variations are observed in the vehicle key updates, which cause fewer credential burdens leading to a high performing V2X. The performance of the system is much affected when the security functions are modeled over mmWave technology as it allows a quick update of sub-divided functions allowing a fast and secure formation of 5G-V2X.

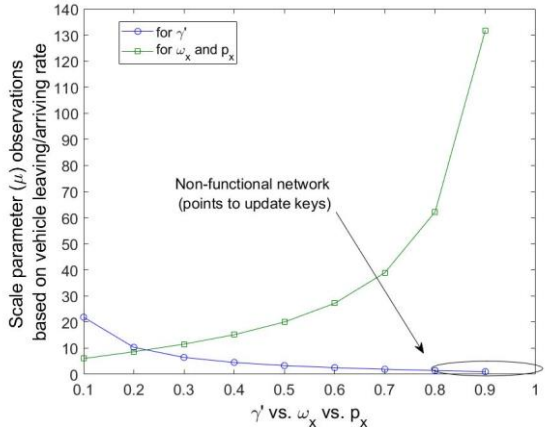

Fig. 10 Scale parameter variations vs. outgoing rate and sustainability governing probability.

The further details of fail-safe durations which can be predicted based on the settings given in Table 2 and by following outcomes from Lemma-4 and Lemma-5 are provided in Table 3. These outcomes show that for the low values of $\mu$, the network cannot be operated with the given settings. However, with a better prediction of the scale parameters, the functional capability of the network increases along with the prediction of fail-safe durations. These results are not fixed; rather these show an observation on how the management of keys can optimize the security of the network while enhancing its sustainability.

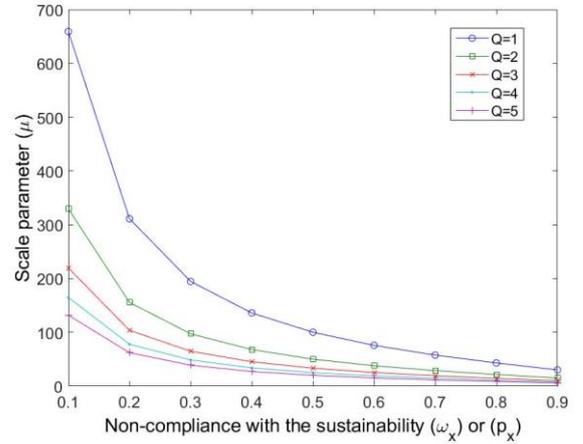

Fig. 11 Scale parameter vs. sustainability governing probability.

It is evident that the sustainability of the network can be enhanced by carefully selecting the authentication protocol which should not cause excessive signaling overheads. Such practice and utilization of the proposed model can provide an optimal solution for managing the security of the backhaul-aware 5G-V2X as proved in this article. In further works, different types of vehicles (such as UAVs [25] [28]) can be considered to test the updates of keys along with the identification of specific-fail safe points. Extremely dynamic vehicles can have severe effects on the system model and the location of servers for supporting 5G security functions. Moreover, the adaptation of the proposed model can be tested in the presence of known attacks, such as host impersonation [29] or emulation attacks [30].

# 8 Conclusions

The V2X formations through 5G enhance the general capabilities of the vehicular network by utilizing the high-performance cellular setups for advanced applications involving real-time media. However, security is always one of the major concerns for the vehicles that utilize cellular services for transmission. To understand such an issue, this article presented a tradeoff between sustainability and the number of key-updates in a backhaul-aware 5G-V2X. A dual security management framework is developed which considers security through long- and short-range authentications and the corresponding sub-divided 5G functions. The proposed solution aims at bringing security near to the users without any impact on the general 5G functions associated with the backhaul operations.

Theoretical and analytical studies are conducted to understand the associated issues with the edge-initiated security in 5G-V2X. In general, the work presented in this article helps to understand the impact of key-updates, joining and leaving rates of vehicles and the number of authentication-passes on the sustainability of the network. Based on these, the proposed dual security management framework identifies multiple fail-safe points up to which the network can be operated without any reconfigurations and risk of cyber-attacks. Multiple sub-functions are derived to facilitate the task of security management. Finally, numerical evaluations are conducted to visually understand the observations of the theoretical and analytical model for obtaining optimal security in the backhaul-aware 5G-V2X.

# 9 Acknowledgment

An initial version of this article was presented at the CISC'W-2018, Sejong University, The Republic of Korea without any publication in the official proceedings. The presented article is referred at [13] and available as arXiv: 1811.08273.

This work was supported by 'The Cross-Ministry Giga KOREA Project' grant funded by the Korea government (MSIT) (No. GK18N0600, Development of 20Gbps P2MP wireless backhaul for 5G convergence service).

## Authors

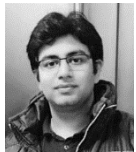

Vishal Sharma received the B.Tech. degree in computer science and engineering from Thapar University, in 2016, and the Ph.D. degree in computer science and engineering from Punjab Technical University, in 2012. He is currently a Research Assi stant Professor in the Department of Information Security Engineering, Soonchunhyang University, South Korea.

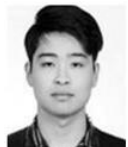

Jiyoon Kim received the B.S. and M.S. degrees in information security engineering from Soonchunhyang University, Asan, South Korea, where he is currently pursuing a Ph.D. degree in the Department of Information Security Engineering.

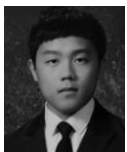

Yongho Ko received a B.S. degree in information security engineering from Soonchunhyang University, Asan, South Korea, where he is currently pursuing a master's degree with the Department of Information Security Engineering.

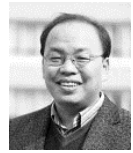

Ilsun You received the M.S. and Ph.D. degrees in computer science from Dankook University, Seoul, South Korea, in 1997 and 2002, respectively, and the Ph.D. degree from Kyushu University, Japan, i n 2012. He is currently an Associate Professor in the Department of Information Security Engineering, Soonchunhyang University, South Korea.

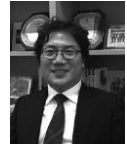

Jung Taek Seo received the Ph.D. degree in information security from the Graduate School of Information Security, Korea University, in 2006. He is currently an Assistant Professor in the Department of Information Security Engineering, Soonchunhyang University, South Korea.